\documentstyle[12pt]{article}
\renewcommand{\baselinestretch}{2.0}
\def \be{\begin{equation}}
\def \ee{\end{equation}}
\def \bea{\begin{eqnarray}}
\def \eea{\end{eqnarray}}
\begin{document}
{\hskip 2 cm } Revised Version  {\hskip 4 cm  IPM-95-110,  hep-th/9507166}
\begin{center}
{\Large{\bf{The Alf'ven Effect and Conformal Field Theory}}}
\vskip 2cm
{\large{M. R. Rahimi Tabar\\and \\ S. Rouhani}}
\vskip 1cm
{\it{Department of Physics, Sharif University of Technology\\Tehran P.O.Box:
11365-9161, Iran.\\Institue for Studies in Theoretical Physics and 
Mathematics
\\ Tehran P.O.Box: 19395-5746, Iran.}}
\end{center}
\vskip 1.5cm
\begin{abstract}
Noting that two-dimensional magnetohydrodynamics can be modeled by conformal
field theory, we argue that when the Alf'ven effect is also taken into
account one is naturally lead to consider conformal field theories, which
have logarithmic terms in their correlation functions. We discuss the
implications of such logarithmic terms in the context of 
magnetohydrodynamics, and derive a relationship between conformal 
dimensions of the velocity stream function, the magnetic flux 
function and the Reynolds number.
\end{abstract}
\newpage
\renewcommand{\baselinestretch}{2.0}
\topmargin -20mm
\oddsidemargin -7mm
\textwidth 165mm
\textheight 230mm
\def \be{\begin{equation}}
\def \ee{\end{equation}}
\def \bea{\begin{eqnarray}}
\def \eea{\end{eqnarray}}
{\bf{1 - Introduction}}

There has been some work on modelling turbulence in two dimensional fluids
by conformal field theory (CFT) [1-7].
Ferretti et al. [6] have generalized Polyakov's method [1] to the case of two
dimensional magnetohydrodynamics (2D - MHD).
We have argued that the existence of a critical dynamical index is
equivalent to the Alf'ven effect [7] i.e. the equipartition of energy between
velocity and magnetic modes [8].
The Alf'ven effect, reduces the number of candidate conformal field theories,
but also it implies that the velocity stream  function $\phi$ and the
magnetic flux function $\psi$ should have similar scaling dimensions.
To reduce the number of candidate conformal field theories other condition
on 2D-MHD, have been imposed by Coceal and Thomas [9].
Gurarie [10] has argued that although in unitary minimal models
two primary fields with the same dimension do not occur , such a situation
can occur in non minimal CFTs.
In such conformal field theories, it has been shown [10] that the correlator
of two fields, has a logarithmic singularity.
\be
<\psi(r) \psi(r^{'})>\sim {|r-r^{'}|}^{-2 h_\psi} \log|r-r^{'}| + \ldots
\ee
Examples of such theories have been studied by Gurarie
[10] , Saleur [11], Rozansky and Saleur in connection with the Wess - Zumino -
Witten model on the super group GL(1,1) [12] and Bilal and Kogan in 
connection with
the gravitational dressed CFT [13,14].
This paper is organised as follows; in section two we give a very brief
summary of magnetohydrodynamics and the Alf'ven effect. In section 3 
we discuss the
implication of the logarithmic divergence and candidate CFT models are given
in section 4 .
\vskip 1.5cm
{\bf{2 - The Alf'ven effect and conformal field theory.}}

The incompressible two dimensional magnetohydrodynamic (2D - MHD) system has
two independent dynamical variables, the velocity stream function
$\phi$ and the
magnetic flux function $\psi$. These obey the pair of equations [15],
\be
{{\partial \omega} \over{\partial t}} = -e_{\alpha \beta} \partial_\alpha \phi
\partial_\beta \omega + e_{\alpha \beta} \partial_\alpha \psi \partial_\beta
J + \mu {\bigtriangledown}^2 \omega
\ee
\be
{{\partial \psi}\over{\partial t}} = -e_{\alpha \beta} \partial_{\alpha}\phi
\partial_\beta \psi + \eta J
\ee
where the vorticity $\omega = \bigtriangledown^2 \phi $ and the current 
$J =
\bigtriangledown^2 \psi$. The two quantitiesy $\mu$ and $\eta$ are the
viscosity
and molecular resistivity, respectively. The velocity and magnetic fields are
given in terms of $\phi$ and $\psi$ :
\be
V_\alpha = e_{\alpha \beta} \partial_\beta \phi
\ee
\be
B_\alpha = e_{\alpha \beta} \partial_\beta \psi
\ee
and $e_{\alpha \beta}$ is the totally antisymmetric tensor, with $e_{12} = 1$.
Chandrasekhar [7] has shown that the Alf'ven effect or the equipartition of
energy between velocity and magnetic modes requires $V^2_k = \alpha B^2_k$, 
with
$\alpha$ of order unity.
In fact he finds $\alpha = 1.62647$ for 2D - MHD. We [8] have argued
that
the existence of a critical dynamical index for 2D - MHD, implies the Alf'ven
effect and if the conformal model holds, this implies the equality of scaling
dimensions of $\phi$ and $\psi$ :
\be
h_\phi = h_\psi
\ee
Here the criteria
of Gurarie [10] are satisfied and these two fields are
logarithmically correlated. 
According to Gurarie [10],
the operator product expansion of two fields $A$ and $B$,
which have two fields
$\phi$ and $\psi$ of equal dimension in their fusion rule [16] , has a 
logarithmic
term:
\be
A(z) B(0) = z^{h_\phi - h_A - h_B}\{\psi(0) + \ldots +\log z(\phi(0) + \ldots )
\}
\ee
to see this it is sufficient to look at four point function :
\be
<A(z_1) B(z_2) A(z_3) B(z_4)> \sim {1\over {(z_1-z_3)}^{h_A}} {1 \over
{(z_2 - z_4)}^{h_B}} {1 \over {[x(1-x)]^{h_A + h_B - h_\phi}}}
F(x)
\ee
Where the cross ratio $x$ is given by :
\be
x = {{(z_1 - z_2)(z_3 - z_4)}\over{(z_1 - z_3)(z_2 - z_4)}}
\ee
In degenerate 
models $F(x)$ satisfies a second order linear differential equation.
Therefore a solution for $F(x)$ can be found in terms of a series expansion :
\be
F(x) = x^\alpha \sum a_n x^n
\ee
In the next section we will show 
that the existence of two fields with equal dimension
in OPE of $A$ and $B$ is equivalent to the secular equation for $\alpha$
having coincident roots, in which case two
independent solutions can be constructed according to :
\be
\sum b_n x^n + \log x \sum a_n x^n
\ee
Now consistency of equation (12) and (8) requires :
\be
<A(z_1) B(z_2) \psi(z_3)> = <A(z_1) B(z_2) \phi(z_3)>\{ \log{(z_1-z_2) \over
{(z_1-z_3)(z_2-z_3)}} + \lambda \}
\ee
\be
<\psi(z) \psi(0)> \sim {1 \over {z^{2h_\psi}}} [\log z +\lambda^{'}]
\ee
\be
<\psi(z) \phi(0)> \sim {1 \over {z^{2 h_\phi}}}
\ee
where $\lambda$ and $\lambda^{'}$ are constants. Eq.(14) is consistent with
the findings of Gurarie [17] and Polyakov [18] which shows that the probability
distribution of such correlation functions is different from the Gibbs
distribution since for the Gibbs distribution we should have
$< \psi(z) \phi(0)> = 0$ . \\
Let us now consider the action of $SL(2,C)$, on the correlator $<\psi(z_1)
\psi(z_2)>$.
The generator of $SL(2,C)$, $(L_0 , L_{\pm 1})$, act on this correlator as
follows;
\bea
&&L_{-1} <\psi(z_1) \psi(z_2)> = 0 \cr
&&L_0  <\psi(z_1) \psi(z_2)> = -2 |z_1 - z_2|^{-2h_\phi} \cr
&&L_{+1}<\psi(z_1) \psi(z_2)> = -2 |z_1 - z_2|^{-2h_\phi} |z_1 + z_2|
\eea

By simple algebra one observes that $<\psi(z_1) \psi(z_2)>$ is 
invariant under$(L_{-1}, L_0^2, L_+ L_0 )$ and also  
$L_0 <\psi(z_1) \psi(z_2)>$ is itself
invariant under the action of $SL(2,C)$.\\ 
Thus $L_0 <\psi(z_1) \psi(z_2)>$
behaves like an ordinary CFT correlation function.
Thus we may solve the resulting first order differential equation for
$<\psi (z_1) \psi(z_2)>$, which naturally leads to a logarithmic singularity.
This result is compatible with the finding in [10] that this type of operator
together with ordinary primary operators form the basis of the Jordan cell
for the operator $L_0$. This fact allows us to find higher-order correlation
functions for the operator $\psi$.
\\
\\
{\bf{3- The Infrared problem and The Energy Spectrum:}}

The presence of logarithmic terms requies a reconsideration of the
infrared problem. The $k$-representation of the correlation is;
\be
<\psi(k) \psi(-k)> = |k|^{-2-2|h_\phi|} [C_1 + \log k]
\ee
which is divergent in the limit of $k\rightarrow 0$ .
One can set some cut-off in the k-space to remove this divergence :
\bea
<\psi(x) \psi(0)> &=& \int^\infty_{k>{1\over R}} k^{-2-2|h_\phi|}
[C + \log k] e^{ik\cdot x} d^2 k \cr
&\sim & R^{2|h_\phi|} (\log R +C^{'}) - x^{2|h_\phi|}(C^{'} + \log X) 
+ \ldots
)
\eea
where $R$ is the large scale of the system.
It seems that it is natural to add some condensate term [1] in momentum space
to cancel the infrared divergence.
The energy spectrum for this type
of correlation, is
\be
E(k) \simeq  k^{-2|h_\phi|+1}(C +\log k)
\ee
which has a logarithmic singularity at the limit of $k\rightarrow 0$.
This spectrum is compatible with the results of Ref. [19] where it has been
shown that, one loop correction to the energy spectrum gives a
logarithmic contribution to the energy spectrum.
\\
\\
{\bf{4- Finding a Candidate Conformal Field Theory.}}

The question is which types of conformal field theory may be 
used for modelling 2D- MHD turbulence, provided we take into account 
the Alf`ven effect as well as the cascade of the mean square
magnetic potential.
At first glance, we cannot use the minimal models.
For a fixed $(p,q)$ all the primary fields in minimal models have 
different dimensions, thus eq.(6) is never satisfied.
To see this let us look at the conformal dimension $h_{m,n}$
of a given primary field $\phi_{m,n}$ 
\be
h_{m,n}={1\over 4 p q} [(mp-nq)^2-(p-q)^2]   
\ee
with $1\leq q\leq p$ ,$ 1\leq m\leq q-1$ and $ 1\leq n\leq p-1$.
Simple algebra shows that if two fields $\phi_{m,n}$ and $\phi_{m',n'}$
have the same dimensions $h_{m,n}= h_{m',n'}$, then we must have 
\be
{m \pm m' \over n \pm n'} = {p\over q}
\ee
And since $p$ and $q$ are coprime, eq.(20) is never satisfied.
However all is not lost, one can find non unitary minimal models  
where two primary fields have almost equal conformal dimensions.
The table of CFT models which are nearly consistent with (6) is given in 
reference [8].
Two primary fields with almost equal conformal dimensions can mimic
logarithmic correlators for a restirected range.\\
According to ref.[20], the hypergeometric equation 
govrning the correlator of
two fields in whose OPE two other fields $ \psi$ and $\phi$ with 
conformal dimesions $h_{\psi}$ and $h_{\psi}+\epsilon $ appear, admits 
two solutions; \\
\bea
&_{2}F_{1}(a,b,c,x)& \\
&x^{\epsilon}\,\, _{2}F_{1}(a+\epsilon,b+\epsilon, c+2\epsilon,x)&
\eea
where $a$, $b$ and $c$ are sums of conformal dimension.
Clearly in the limit of $\epsilon \rightarrow 0 $ these two solutions 
coincide. 
Another independent solution exists, it involves logarithms 
and can be generated by standard methods [21]. 
Therefore expanding the above solutions neae $\epsilon=0$
logarithmic behaviour is obtained. \\ 
Construct two fields $\Phi_{+}$ and $\Phi_{-}$,
\be
\Phi_{\pm}= \psi \pm i\lambda \phi
\ee
where $\lambda$ is a constant with dimension $\epsilon$.
Then the correlators are 
\be 
<\Phi_{+}(z) \Phi_{+}(0)> \simeq z^{-2 h_{\psi}} \ln z 
\ee
and 
\be
<\Phi_{+}(z) \Phi_{-}(0)> \simeq z^{-2 h_{\psi}}
\ee
provided $z$ lies in the range:
\be
a \ll z \ll R
\ee
where $a$ is the dissipation range, this gives:
\be
\lambda \simeq R^{\epsilon} \hskip 1.5cm \epsilon 
\leq {1 \over {5/2 \ln R_{e}}}
\ee 
where $ R_{e} $ is the typical Reynold`s number of system and 
we have used the 
relation  $ a \simeq R  {R_{e}}^{-5/2}$, which can easily be seen using 
dimensional arguments.
 For example turbulence of up to $R_{e} \sim 10^{12}$ may be describ
 by the minimal model $(6,35)$, which has $\epsilon= 1/70$. \\
The above is of course an approximate argument
and the approximation improves as $\epsilon$ tends towards zero or
the central charge tends towards unity.
If we insist on exact
logarithmic correlators we need to consider other CFT 's , probably
with the effective central charge equal to unity [22] and also see [23].
Work in this direction is under progress. \\
{\bf{Acknowledgements}}
The authors are grateful to S. Parvizi for stimulating discussions, and
wish to thank Werner Nahm and the referee for pointing out an error 
in the earlier version of
this article.
\\

\end{document}